\documentclass[conference]{IEEEtran}
\IEEEoverridecommandlockouts
\usepackage{cite}
\usepackage{amsmath,amssymb,amsfonts}
\usepackage[ruled]{algorithm2e}
\DontPrintSemicolon
\usepackage{graphicx}
\usepackage{textcomp}
\usepackage{xcolor}
\usepackage{booktabs}
\usepackage{multirow}
\usepackage[hidelinks]{hyperref}
\usepackage{makecell}
\def\BibTeX{{\rm B\kern-.05em{\sc i\kern-.025em b}\kern-.08em
    T\kern-.1667em\lower.7ex\hbox{E}\kern-.125emX}}

\makeatletter
\newcommand{\linebreakand}{%
\end{@IEEEauthorhalign}
\hfill\mbox{}\par
\mbox{}\hfill\begin{@IEEEauthorhalign}
}
\makeatother

\begin{document}

\title{Attention Weighted Mixture of Experts with Contrastive Learning for Personalized Ranking in E-commerce}

\author{\IEEEauthorblockN{Juan Gong}
\IEEEauthorblockA{\textit{JD.com} \\
Beijing, China \\
gongjuan199@163.com}
\and
\IEEEauthorblockN{Zhenlin Chen}
\IEEEauthorblockA{\textit{JD.com} \\
Beijing, China \\
chenzhenlin6@jd.com}
\and
\IEEEauthorblockN{Chaoyi Ma}
\IEEEauthorblockA{\textit{JD.com} \\
Beijing, China \\
machaoyi1227@163.com}
\and
\IEEEauthorblockN{Zhuojian Xiao}
\IEEEauthorblockA{\textit{JD.com} \\
Beijing, China \\
xiaozhuojian5@jd.com}
\and
\IEEEauthorblockN{Haonan Wang}
\IEEEauthorblockA{\textit{JD.com} \\
Beijing, China \\
wanghaonan@jd.com}
\linebreakand
\IEEEauthorblockN{Guoyu Tang}
\IEEEauthorblockA{\textit{JD.com} \\
Beijing, China \\
tangguoyu@jd.com}
\and
\IEEEauthorblockN{Lin Liu}
\IEEEauthorblockA{\textit{JD.com} \\
Beijing, China \\
liulin1@jd.com}
\and
\IEEEauthorblockN{Sulong Xu}
\IEEEauthorblockA{\textit{JD.com} \\
Beijing, China \\
xusulong@jd.com}
\and
\IEEEauthorblockN{Bo Long}
\IEEEauthorblockA{\textit{JD.com} \\
Beijing, China \\
bo.long@jd.com}
\and
\IEEEauthorblockN{Yunjiang Jiang}
\IEEEauthorblockA{\textit{JD.com} \\
Mountain View, CA, USA \\
yunjiangster@gmail.com}
}

\maketitle

\begin{abstract}
Ranking model plays an essential role in e-commerce search and recommendation. An effective ranking model should give a personalized ranking list for each user according to the user preference. Existing algorithms usually extract a user representation vector from the user behavior sequence, then feed the vector into a feed-forward network (FFN) together with other features for feature interactions, and finally produce a personalized ranking score. Despite tremendous progress in the past, there is still room for improvement. Firstly, the personalized patterns of feature interactions for different users are not explicitly modeled. Secondly, most of existing algorithms have poor personalized ranking results for long-tail users with few historical behaviors due to the data sparsity.

To overcome the two challenges, we propose Attention Weighted Mixture of Experts (AW-MoE) with contrastive learning for personalized ranking. Firstly, AW-MoE leverages the MoE framework to capture personalized feature interactions for different users. To model the user preference, the user behavior sequence is simultaneously fed into expert networks and the gate network. Within the gate network, one gate unit and one activation unit are designed to adaptively learn the fine-grained activation vector for experts using an attention mechanism. Secondly, a random masking strategy is applied to the user behavior sequence to simulate long-tail users, and an auxiliary contrastive loss is imposed to the output of the gate network to improve the model generalization for these users. This is validated by a higher performance gain on the long-tail user test set.

Experiment results on a JD real production dataset and a public dataset demonstrate the effectiveness of AW-MoE, which significantly outperforms state-of-art methods. Notably, AW-MoE has been successfully deployed in the JD e-commerce search engine, serving the real traffic of hundreds of millions of active users.
\end{abstract}


\section{Introduction}\label{introduction}
In the past decades, increasingly people have got used to shopping on e-commerce websites (e.g., Amazon, eBay, Alibaba, and JD), where they find interesting products either by issuing a specific query or exploring products recommended by the websites. At the core of the success of an e-commerce website is its search engine and recommendation system. A well-designed search engine or recommendation system, which gives a personalized product list for each user according to the user preference, not only increases revenues for websites but also brings convenient online shopping experiences for consumers.

\begin{figure*}[t]
	\centering
	\includegraphics[width=1.0\linewidth]{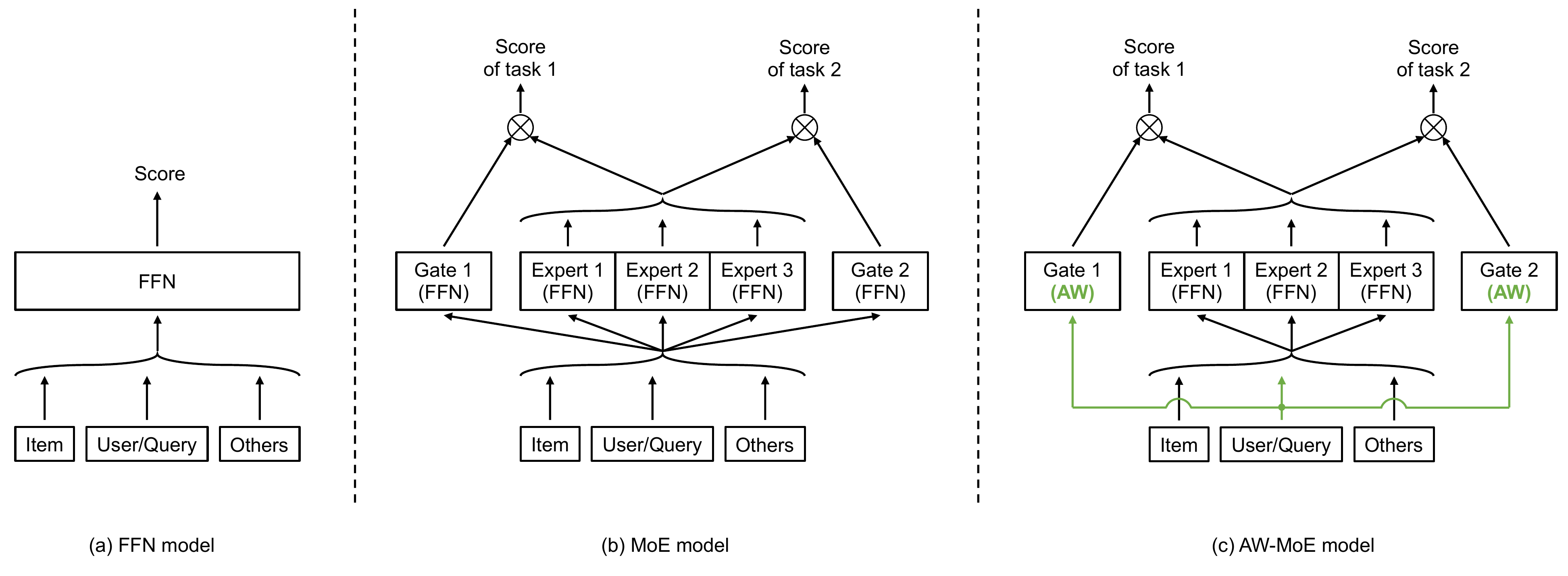}
	\caption{Comparison between existing models and our proposed AW-MoE model.}
	\label{model-comparison}
\end{figure*}

The effectiveness of a search engine or a recommendation system critically relies on the ranking model, which should promote relevant products and suppress the irrelevant ones according to users' preferences. Existing ranking models~\cite{covington2016deep-YouTube-DNN,zhou2018deep-DIN,zhou2019deep-DIEN,feng2019deep-DSIN,pi2019practice-MIMN,chen2019behavior-BST} usually extract a user representation vector from the user behavior sequence, then feed the vector into a feed-forward network (FFN) together with other features for feature interactions, and finally produce a personalized ranking score, which is usually the Click Through Rate (CTR) or the Conversion Rate (CVR) (Figure  \ref{model-comparison}a).

Although encouraging progress has been made, building a personalized ranking model still remains challenging due to a huge number of online users whose interests are highly diverse. Firstly, the personalized patterns of feature interactions for different users are not explicitly modeled. Most models have only one FFN, which limits them to learn diverse patterns of feature interactions for different users (Figure \ref{model-comparison}a). Although some works~\cite{zhao2019recommending-YouTube-MMoE,gu2020deep-DMT-MMoE} design multiple FFNs in their ranking models by adopting the mixture of experts (MoE) framework~\cite{shazeer2017outrageously-Top-K-MoE}, their gate networks are vanilla FFNs and are mainly responsible for learning experts activation according to different tasks instead of different users, which we call the task-oriented MoE (Figure \ref{model-comparison}b). In the task-oriented MoE models, multiple experts are still shared by all users, and the personalized patterns of experts activation for different users are not explicitly modeled.

\begin{figure}[h]
	\centering
	\includegraphics[width=0.75\linewidth]{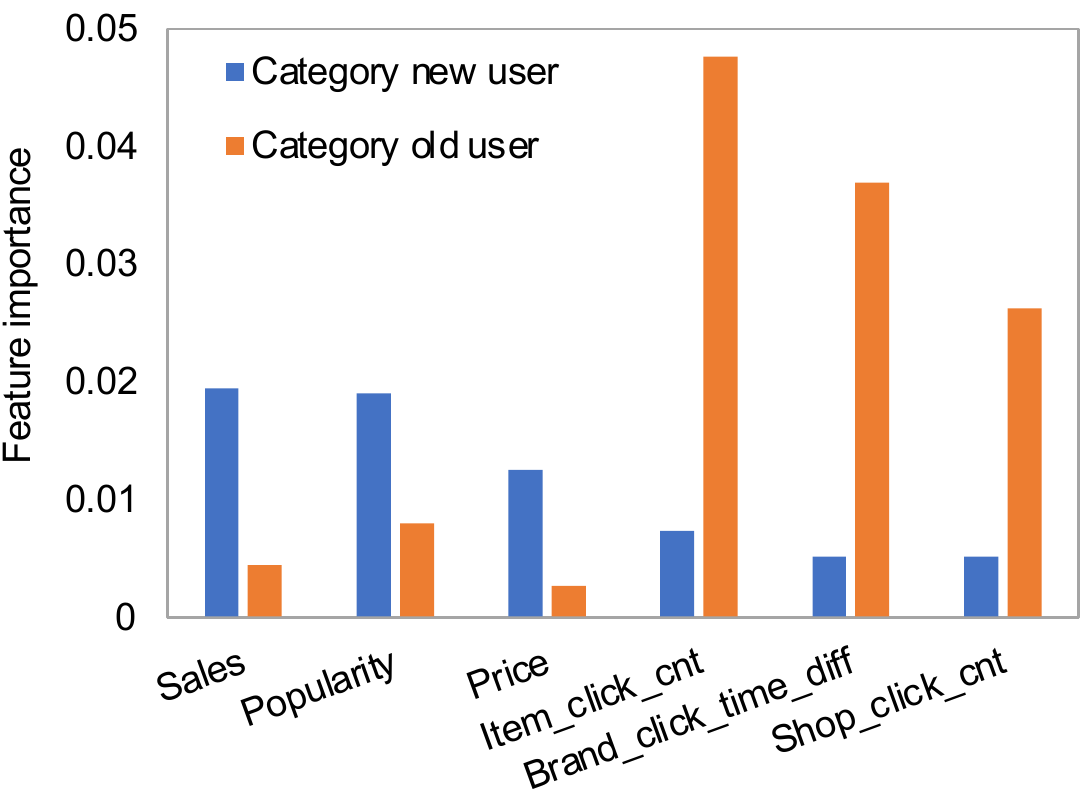}
	\caption{Feature importance calculated by XGBoost~\cite{chen2016xgboost} for different user groups.}
	\label{feature-importance-distribution}
\end{figure}

For example, we find that different users often pay attention to different aspects of input features. Figure \ref{feature-importance-distribution} shows that features related to the popularity of the target item, such as sales volume and price, are much more important for users who have no historical behaviors in the category of the target item (defined as \textbf{category new users}) than for users who have (\textbf{category old users}). On the contrary, two-sided features between the user and the target item, such as click counts for the item or the shop and the click time difference for the brand of the item, are much more important for category old users than for category new users. This observation is expected because category new users have no prior knowledge for the target item or category, they tend to follow the general trend or are affected by the price of the item, whereas category old users have well-formed shopping habits in the category and they are more likely to purchase items according to their preferences. This finding motivates us to explicitly model personalized feature interactions for different users, which is not the case in existing algorithms.

Secondly, obstacles such as data sparsity stand in the way of applying current ranking models to long-tail users who have limited historical behaviors. Several methods have been proposed to address the long-tail user problem. Some methods leverage various user profile information, such as demographic data (e.g., gender, age, location, etc.) ~\cite{volkovs2017-DropoutNet,liang2020joint-JTCN}, and social relationships~\cite{sedhain2014social-FB} to generate appropriate ranking lists for long-tail users. Some methods utilize data from external domains as prior knowledge to infer preferences of long-tail users in the target domain~\cite{Zang2022-CDR,li2020-Ddtcdr}. Although these two types of methods demonstrate their effectiveness, they are facing data privacy issues. Other methods such as meta-learning~\cite{lee2019-Melu,dong2020-Mamo} and generative adversarial networks~\cite{chen2022-Coldgan,chen2022-GAR}  show promising results, but require substantial modifications to the current ranking system, which is too expensive in industry.

Intuitively, users with similar historical behaviors have similar shopping interests. Although long-tail users have limited historical behaviors, their interests can be learned from ordinary users with similar but more abundant historical behaviors. For example, user A is an ordinary user and has bought many running products, such as sneakers, quick-dry shirts, and sports watches; while user B is a long-tail user and has bought only one pair of sneakers. When user B searches for "watches", it is more reasonable to rank sports watches higher than business watches if the ranking model has learned the latent similarity between user A and user B. This intuition motivates us to guide the model to learn similarities between long-tail users and ordinary users when training.

To overcome the two aforementioned challenges, we propose Attention Weighted Mixture of Experts (AW-MoE) with contrastive learning for effective personalized ranking. AW-MoE adopts the MoE~\cite{shazeer2017outrageously-Top-K-MoE} framework for personalized ranking, within which multiple expert networks are used to model the diverse feature interaction patterns, and the gate network is used to explicitly learn the personalized experts activation vectors for different users (Figure \ref{model-comparison}c), hence we call AW-MoE the user-oriented MoE. To capture the fine-grained user preference for experts activation, the user behavior sequence is simultaneously fed into expert networks and the gate network. Within the gate network, one gate unit and one activation unit are designed for each item in the user behavior sequence, and the final output of the gate network is computed as the weighted sum of outputs from all gate units using an attention mechanism. AW-MoE is expected to be more effective in selecting personalized experts according to the user preference, expressed through the attention weighted gate network.

To alleviate data sparsity of long-tail users, items in the user behavior sequence are randomly masked during network training to simulate long-tail users, and an auxiliary contrastive loss is imposed to the output of the gate network to guide the model to learn similarities between long-tail users and ordinary users. The elaborately designed gate network together with the contrastive loss contribute to significant performance gain in personalized ranking tasks for both ordinary users and long-tail users.

In summary, we make the following contributions:

\begin{itemize}
	\item We propose AW-MoE, a novel model for personalized ranking based on the MoE framework. The key idea is to model the diverse feature interaction patterns by multiple experts, and to explicitly learn the personalized experts activation vectors by the attention weighted gate network.
	
	\item We propose a contrastive learning strategy to guide the model to learn similarities between long-tail users and ordinary users, and hence improve the model generalization for long-tail users.
	
	\item We validate the effectiveness of AW-MoE on both search and recommendation datasets. Experimental results show that our proposed model outperforms existing state-of-the-art methods for both ordinary users and long-tail users. Besides, different user groups have been found to activate different experts, which improves the learning effectiveness for each expert. AW-MoE has now been deployed in the JD e-commerce search engine, serving the real traffic of hundreds of millions of active users.
\end{itemize}

\section{RELATED WORK}

Our work introduces the user behavior sequence modeling and  contrastive learning to the framework of mixture of experts. Thus, we review recent progress in user behavior sequence modeling, mixture of experts, and contrastive learning for references.

\subsection{User Behavior Sequence Modeling}

User behavior sequence modeling, which aims to mine a user's interests from a sequence of items interacted by the user, is an important module in both search engine and recommendation system. As a pioneering work, Youtube DNN aggregates embeddings of all items into a fixed-length vector by sum-pooling to represent user interests~\cite{covington2016deep-YouTube-DNN}. Later on, some algorithms, such as DIN~\cite{zhou2018deep-DIN} and DSTN~\cite{ouyang2019deep-DSTN}, use the attention~\cite{bahdanau2014neural-Attention} between the target item and items in the user behavior sequence to adaptively learn the representation vector of user interests. Since algorithms using pooling or attention take the user behavior sequence as an unordered set of items, they fail to capture the sequential nature of the user behavior sequence. To address this limitation, Recurrent Neural Network (RNN)-based algorithms like GRU4Rec~\cite{hidasi2015session-GRU4Rec} and DIEN~\cite{zhou2019deep-DIEN} are proposed. Recent works, such as BERT4Rec~\cite{sun2019bert4rec-BERT4Rec}, SIM~\cite{pi2020search-SIM}, and KFAtt~\cite{liu2020kalman-KFAtt}, make significant progress by applying the self-attention mechanism~\cite{vaswani2017attention-transformer} in the user behavior sequence modeling. Although the user behavior sequence is modeled by different methods in the aforementioned algorithms, it is exclusively used as an input feature for user representation modeling, and is used only once (Figure \ref{model-comparison}ab). On the contrary, our proposed AW-MoE fully exploits the user behavior sequence twice, once for user representation modeling in expert networks similar to existing algorithms, and again for experts activation by feeding it into the gate network (Figure \ref{model-comparison}c).

\subsection{Mixture of Experts}

Ensemble learning is a ubiquitous technique in machine learning that reliably boosts performance of constituent models or signals~\cite{sagi2018ensemble}. One disadvantage of putting multiple models in an ensemble is its added training and serving cost. Mixture of Experts (MoE), first introduced in~\cite{jacobs1991adaptive-MoE} and re-invented in~\cite{shazeer2017outrageously-Top-K-MoE} under the neural network context, reduces training and inference time at the expense of extra model parameters. The idea of MoE is to partition a problem into several subspaces, each of which is learned by an expert network, and a gate network is designed to aggregate results from all experts. As a follow-up, Multi-gate Mixture of Experts (MMoE) applies MoE to multi-task learning, and for each task allocates a gate network to learn the task-specific functionality~\cite{ma2018modeling-MMoE}. As a building block, MoE has been successfully applied in the field of Computer Vision (CV)~\cite{wang2020deep-MoE-in-CV1,riquelme2021scaling-MoE-in-CV2} and Natural Language Processing (NLP)~\cite{lepikhin2020gshard-MoE-in-NLP1,fedus2021switch-MoE-in-NLP2}. The applications of MoE in search engine and recommendation system mainly focus on multi-objective optimization~\cite{zhao2019recommending-YouTube-MMoE,gu2020deep-DMT-MMoE} (Figure \ref{model-comparison}b). For example, the YouTube recommendation system uses MoE to simultaneously optimize the user engagement objective and the user satisfaction objective~\cite{zhao2019recommending-YouTube-MMoE}, and the JD recommendation system uses MoE to simultaneously optimize tasks of CTR prediction and CVR prediction~\cite{gu2020deep-DMT-MMoE}. An exception is~\cite{xiao2021adversarial-HSC-MoE}, which optimizes only one objective but uses MoE to learn a ranking model that specializes for each query category by feeding the category extracted from the user query into the gate network. In light of~\cite{xiao2021adversarial-HSC-MoE}, we propose AW-MoE, which uses MoE to learn a ranking model that specializes for each user by feeding the user behavior sequence into the gate network (Figure \ref{model-comparison}c).

\subsection{Contrastive Learning}

Recently, contrastive learning methods~\cite{le2020contrastive-CL-review} have achieved remarkable successes on learning useful representations in a self-supervised manner. Their basic idea is to maximize similarity between positive example pairs and to minimize similarity between negative example pairs. Most contrastive learning works focus on exploiting data augmentation to design self-supervised learning pretext tasks, which are closely related to the data domains. In the CV field, data augmentations include colorization~\cite{zhang2016colorful}, rotation~\cite{gidaris2018unsupervised}, and cropping~\cite{chen2020simple}; and in the NLP field,  data augmentations include term masking~\cite{devlin-etal-2019-bert} and sentence reordering~\cite{lewis2020bart}. The contrastive learning framework has also been applied to ranking models of search engine and recommendation system. Some works design item-level augmentations, such as masking items or its attributes~\cite{zhou2020s3,yao2021self}, to improve item representation learning; while others design user-level augmentations, such as masking or reordering user behavior sequences~\cite{xie2022contrastive,zhu2021contrastive}, to improve user representation learning. Inspired by these works, we exploit the contrastive learning framework  to improve the user representation learning of our AW-MoE model. During training, we randomly mask the user behavior sequence and impose a contrastive loss to the output of the gate network, which is regarded as the user representation. After random masking, the augmented user behavior sequences simulate long-tail users, thus model generalization for these users is enhanced.

\section{THE RANKING MODEL IN JD E-COMMERCE SEARCH ENGINE}

Although the proposed AW-MoE is applicable to ranking models in both the search engine and the recommendation system, we mainly focus on the search scenario in this section, and show the applicability to the recommendation scenario in section \ref{experimental-results-section}. We first give an overview of the architecture of the ranking model in subsection \ref{the-overall-architecture-subsection}. Subsection \ref{the-input-network-subsection} describes the input network, which is shared with some of our baseline models such as DIN~\cite{zhou2018deep-DIN}. Subsection \ref{the-attention-weighted-moe-subsection} goes into the detailed design of our novel AW-MoE model architecture. Subsection \ref{contrastive-learning-subsection} introduces the contrastive learning strategy when training our model. Subsection \ref{algorithm-analysisn-subsection} summarizes AW-MoE as an algorithm and analyzes the time and space complexities. Lastly, we discuss practical strategies for system deployment in subsection \ref{system-deployment-subsection}.

\subsection{The Overall Architecture}\label{the-overall-architecture-subsection}

The goal of a ranking model is to predict the probability that a $user$ interacts with an $item$ under some $contexts$. For the sake of simplicity, this scenario is denoted as an impression $(user, item, context)$~\cite{liu2020category-CSCNN}. In the e-commerce search scenario, features of an impression consist of user behavior sequence, target item, query, and other features. Specifically, other features consist of user profile information, and cross features between the user and the item.

For model training, a dataset $D = \{(\boldsymbol{x}_1,y_1), ..., (\boldsymbol{x}_{N},y_{N})\}$ is collected from the search log, and $\boldsymbol{x}_i\in \mathbb{R}^d$ is the overall feature vector of dimension $d$ for the $i$-th impression, $y_i\in\{0,1\}$ is the class label indicating a click for CTR prediction or a purchase for CVR prediction, and $N$ is the number of training examples. The training objective is to minimize the negative log-likelihood loss defined as:
\begin{equation}
	\mathcal{L}_{rank}=-\frac{1}{N}\sum_{i=1}^N(y_i\log(\hat{y_i})+(1-y_i)\log(1-\hat{y_i})),
\end{equation}
where $\hat{y_i}$ is the output of the ranking model, representing the predicted CTR or CVR.

\begin{figure*}[h]
	\centering
	\includegraphics[width=\linewidth]{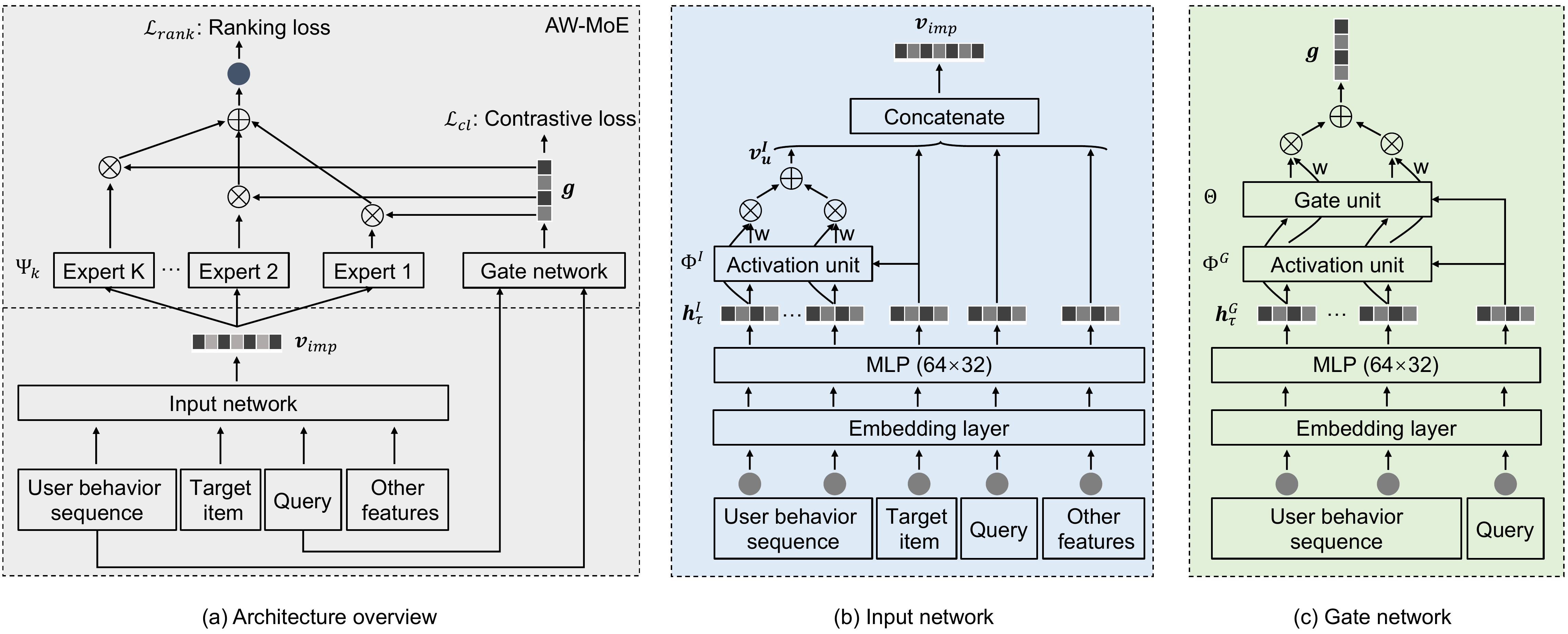}
	\caption{The architecture of our ranking model: (a) architecture overview, (b) architecture of the input network, (c) architecture of the gate network.}
	\label{the-overall-architecture-of-the-ranking-model}
\end{figure*}

As shown in Figure \ref{the-overall-architecture-of-the-ranking-model}a, the overall architecture of the ranking model consists of two components, the input network and the AW-MoE. First, all raw features of an impression are fed into the input network to learn the feature representation of the impression. Then, the output of the input network is fed into the AW-MoE for feature interaction, which is guided by the user behavior sequence and the query. At last, AW-MoE outputs the prediction of CTR or CVR for the input impression. We elaborate the two components in the following subsections.

\subsection{The Input Network}\label{the-input-network-subsection}

As shown in Figure \ref{the-overall-architecture-of-the-ranking-model}b, the input network is used to learn the representation vector of an input impression. First, all raw features of an impression are transformed to vectors using an embedding layer. We denote the embedding vectors of the $j$-th item in the user behavior sequence, target item, query, and other features as $\boldsymbol{e}_{b_j}$, $\boldsymbol{e}_t$, $\boldsymbol{e}_q$, and $\boldsymbol{e}_o$, respectively. Then, all embedding vectors are fed into a multi-layer perceptron (MLP) to get the hidden vectors:
\begin{equation}\label{eq2}
	\boldsymbol{h}_\tau^I={\rm MLP}^I(\boldsymbol{e}_\tau), \tau\in\{b_j,t,q,o\},
\end{equation}
where $\boldsymbol{h}_\tau^I$ is the hidden vector for feature type $\tau$, and the superscript $I$ indicates variables or functions in the input network.

\begin{figure}[h]
	\centering
	\includegraphics[width=\linewidth]{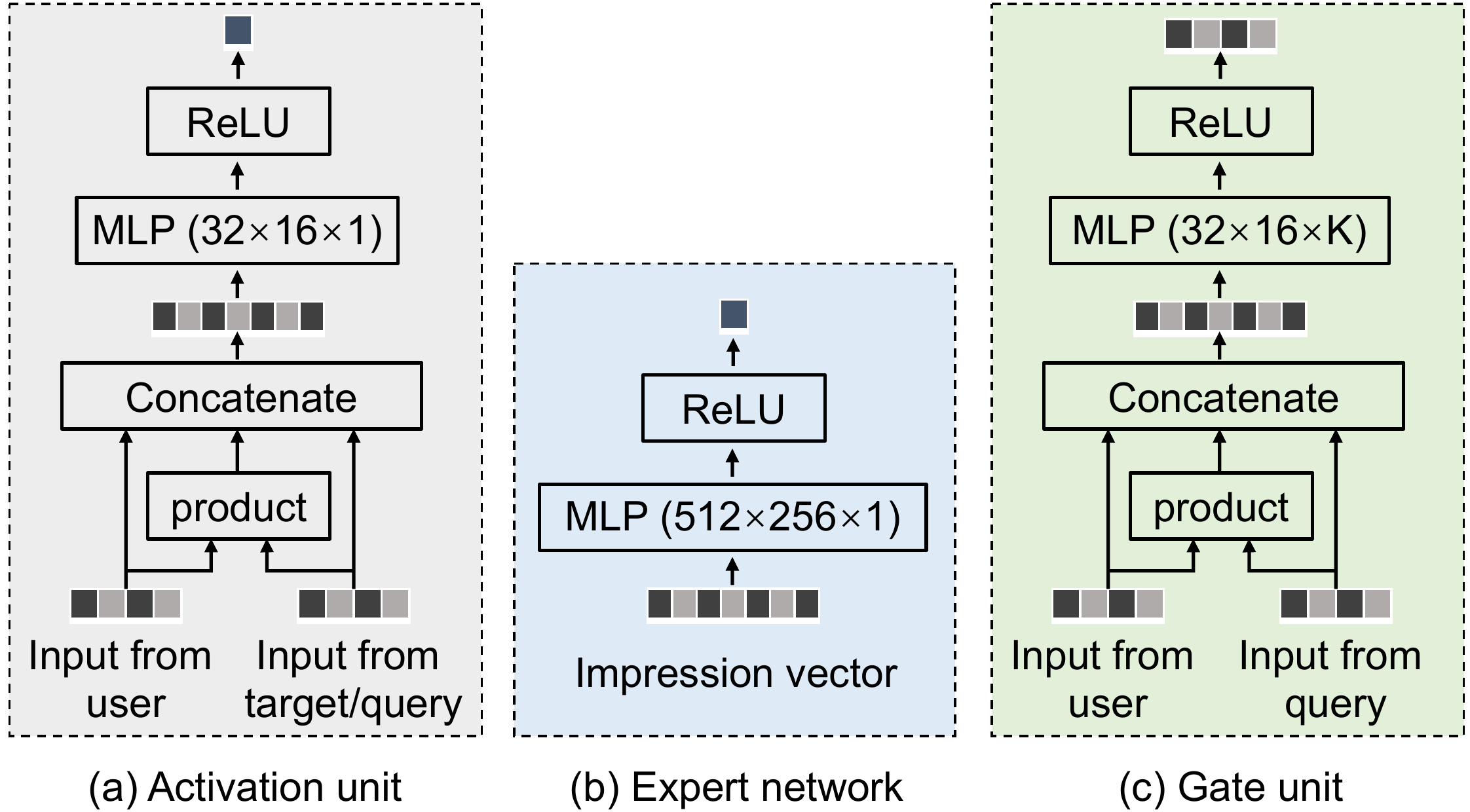}
	\caption{Modules of our proposed attention weighted MoE: (a) the activation unit, (b) the expert network, (c) the gate unit, where $K$ is the number of expert networks.}
	\label{the-architecture-of-activation-gate-expert}
\end{figure}

Next, the user representation vector $\boldsymbol{v}_u^I$ given a target item $t$ is learned by a DIN~\cite{zhou2018deep-DIN} architecture using the attention weighted sum of hidden vectors of all items in the user behavior sequence, as shown in Eq.(\ref{eq3})
\begin{equation}\label{eq3}
	\boldsymbol{v}_u^I=\sum_{j=1}^M\Phi^I(\boldsymbol{h}_{b_j}^I,\boldsymbol{h}_t^I)\boldsymbol{h}_{b_j}^I=\sum_{j=1}^Mw_j^I\boldsymbol{h}_{b_j}^I,
\end{equation}
where $\Phi^I(\cdot)$ is the activation unit, an FFN with output $w_j^I\in\mathbb{R}$ as the attention score, as illustrated in Figure \ref{the-architecture-of-activation-gate-expert}a, and $M$ is the length of the user behavior sequence. In this way, different items in the user behavior sequence are activated when encountering different target items, leading to a user representation vector specific to the target item.

At last, the impression representation vector is obtained by concatenating vectors of user, target item, query, and other features, as shown in Eq.(\ref{eq4}) 
\begin{equation}\label{eq4}
	\boldsymbol{v}_{imp}=\boldsymbol{v}_u^I||\boldsymbol{h}_t^I||\boldsymbol{h}_q^I||\boldsymbol{h}_o^I,
\end{equation}
where $||$ indicates the concatenation operation.

\subsection{The Attention Weighted MoE}\label{the-attention-weighted-moe-subsection}

Once the impression representation vector is obtained, it is usually fed into an FFN for feature interactions in conventional ranking models. In this way, all input users share a same FFN for feature interactions. However, different users often pay attention to different aspects of input features and the patterns of feature interactions vary over users, the ability to learn different patterns of feature interactions for a considerable number of online users is limited by only one shared FFN.

We address this issue by proposing a novel MoE specifically for personalized ranking, Attention Weighted MoE, which learns multiple expert networks for feature interactions and exploits the user behavior sequence to activate experts specific to the user using an attention mechanism.

\subsubsection{The Expert Network}

For each expert network $k$, it takes the impression representation vector as the input, and outputs a ranking score $s_k\in \mathbb{R}$, as shown in Eq.(\ref{eq5}).
\begin{equation}\label{eq5}
	s_k=\Psi_k(\boldsymbol{v}_{imp}),
\end{equation}
where $\Psi_k (\cdot)$ is the $k$-th expert network. Please note that all experts share an identical network structure, an FFN as illustrated in Figure \ref{the-architecture-of-activation-gate-expert}b, but have different parameters due to random initialization.

\subsubsection{The Gate Network}

The gate network is designed to learn an activation vector specific to the current user in the current session, and the learned activation vector will be used for weighted sum of ranking scores from all expert networks. For personalization, the gate network takes the user behavior sequence and the query as inputs, and outputs a vector $\boldsymbol{g}\in\mathbb{R}^K$, where $K$ is the number of expert networks and $g_k\in\mathbb{R}$ is the activation score for the $k$-th expert. 

As $\boldsymbol{g}$ is learned from the user behavior sequence, it can also be regarded as a user representation similar to $\boldsymbol{v}_u^I$ in the input network. Their difference lies in that $\boldsymbol{v}_u^I$ is for feature interactions, and $\boldsymbol{g}$ is for experts activation.

As shown in Figure \ref{the-overall-architecture-of-the-ranking-model}c, the gate network first transforms items in the user behavior sequence and the query to vectors using the embedding layer same as that in the input network. Then, all embedding vectors are fed into an MLP to get the hidden vectors:
\begin{equation}
	\boldsymbol{h}_\tau^G={\rm MLP}^G(\boldsymbol{e}_\tau), \tau\in\{b_j,q\},
\end{equation}
where $\boldsymbol{h}_\tau^G$ is the hidden vector for feature type $\tau$, and the superscript $G$ indicates variables or functions in the gate network. Please note that the MLP in the gate network does not share parameters with that in the input network, hence the hidden vectors in the gate network are not the same as those in the input network.

Next, for each item $j$ in the user behavior sequence, a gate unit is designed to learn an activation vector $\boldsymbol{a}_j\in\mathbb{R}^K$:
\begin{equation}
	\boldsymbol{a}_j=\Theta(\boldsymbol{h}_{b_j}^G,\boldsymbol{h}_q^G),
\end{equation}
where $\Theta(\cdot)$ is the gate unit, an FFN as illustrated in Figure \ref{the-architecture-of-activation-gate-expert}c, and $a_{jk}\in\mathbb{R}$ is the activation score for the $k$-th expert from the $j$-th item. With the gate unit, each item in the user behavior sequence learns a vector for experts activation, thus capturing the fine-grained user preference. Please note that the network structure of the gate unit is similar to that of the activation unit except that the output of the gate unit is a vector while that of the activation unit is a scalar.

At last, for each expert network $k$, the final activation score $g_k\in\mathbb{R}$ is calculated by weighted sum of $k$-th scores from all gate units:
\begin{equation}
	g_k=\sum_{j=1}^M\Phi^G(\boldsymbol{h}_{b_j}^G,\boldsymbol{h}_q^G)a_{jk}=\sum_{j=1}^Mw_j^Ga_{jk},
\end{equation}
where $\Phi^G(\cdot)$ is the activation unit with output $w_j^G\in\mathbb{R}$ as the attention score in the gate network, and it shares an identical network structure with $\Phi^I(\cdot)$ but has different parameters due to random initialization.

In summary, for each item in the user behavior sequence, the gate unit is designed to learn the activation scores for all experts, and the activation unit is designed to learn the attention score of the item. Therefore, we denote the novel ranking model as the Attention Weighted MoE (AW-MoE). With AW-MoE, a fine-grained activation vector $\boldsymbol{g}\in\mathbb{R}^K$ is learned from all items in the user behavior sequence using an attention mechanism. 

Please note that although both the input network and the gate network have a module for user behavior sequence modeling, their motivations are different. For the input network, the learned user representation $\boldsymbol{v}_u^I$, together with other features are fed into multiple expert networks for feature interactions, which is the common case in existing algorithms; while for the gate network, the learned user representation, the output of the gate network $\boldsymbol{g}$, is responsible for experts activation. Furthermore, in the gate network, one extra gate unit is designed for each item in the user behavior sequence to learn the fine-grained activation of the gate network to expert networks. To our best knowledge, this is the first work that the user behavior sequence has been simultaneously fed into expert networks and the gate network for personalized ranking.

Once the outputs of all expert networks and the gate network are obtained, the final ranking score of the AW-MoE is calculated as the weighted sum of the outputs from all experts:
\begin{equation}
	\hat{y}=\sum_{k=1}^Kg_ks_k.
\end{equation}

\subsection{Contrastive Learning}\label{contrastive-learning-subsection}

According to the design of the AW-MoE network, the output of the gate network is also regarded as the user representation, which is further used to activate different experts. In order to improve the model generalization for long-tail users, the input user behavior sequences are randomly masked to simulate long-tail users, and an auxiliary contrastive loss is imposed to outputs of the gate network to pull user representations of the masked and the original user behavior sequences close. In so doing, long-tail users learn from ordinary users, hence data sparsity of long-tail users is alleviated and the model generalization for these users is enhanced. 

Moreover, the contrastive learning strategy brings an extra benefit for the robustness of the model. As a user's behavior is flexible, one can interact with different items for the same shopping need, the observed user behavior sequence is not definitive and exact. The randomly masked user behavior sequence and the contrastive loss force the model to be more robust to possible variations of user behavior sequences.

\begin{figure}[h]
	\centering
	\includegraphics[width=\linewidth]{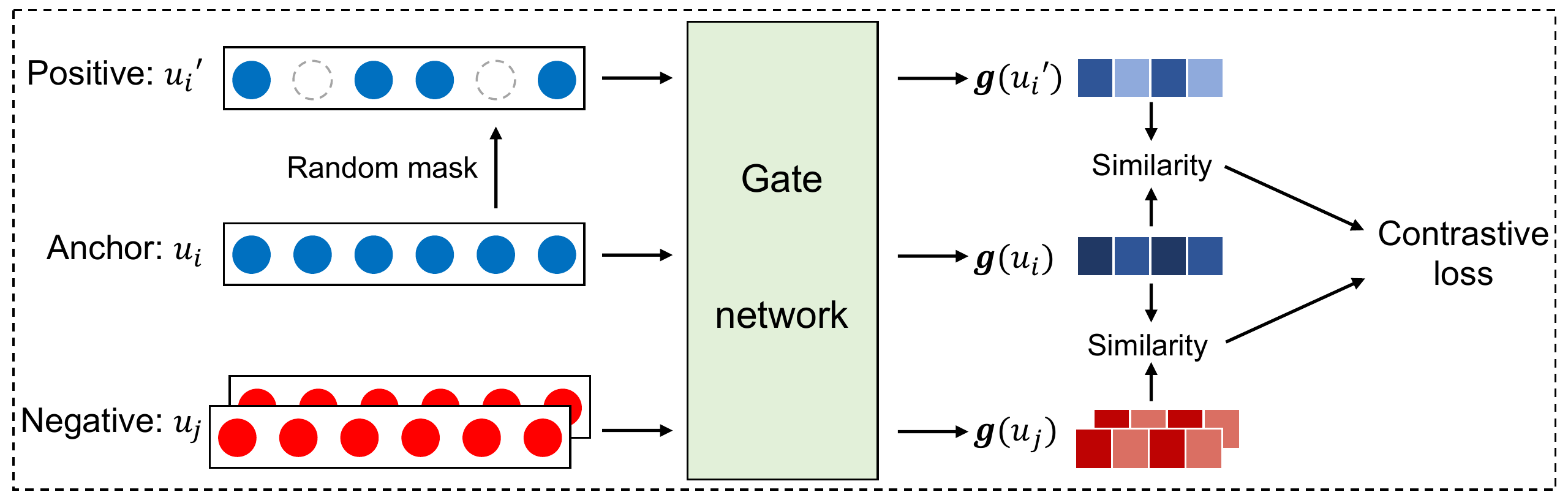}
	\caption{The contrastive learning strategy.}
	\label{the-contrastive-learning}
\end{figure}

Specifically, as shown in Figure \ref{the-contrastive-learning}, for each user $u_i$, one positive user instance $u_i^{\prime}$ is generated by randomly masking items in the user behavior sequence of $u_i$ with probability $p$, and $l$ negative user instances $u_j$ are randomly sampled from in-batch training examples. All of $u_i$, $u_i^{\prime}$ and  $u_j$  are fed into the gate network, and their user representations, $\boldsymbol{g}(u_i)$, $\boldsymbol{g}(u_i^{\prime})$ and $\boldsymbol{g}(u_j)$, are obtained, respectively. Finally, the infoNCE loss~\cite{oord2018representation-infoNCE} is calculated as the contrastive learning loss:


\begin{equation}
	\mathcal{L}_{cl}=-\text{log}\frac{\text{exp}(f(\boldsymbol{g}(u_i),\boldsymbol{g}(u_i^{\prime})))}{\text{exp}(f(\boldsymbol{g}(u_i),\boldsymbol{g}(u_i^{\prime})))+\sum\limits_{j=1}^{l}\text{exp}(f(\boldsymbol{g}(u_i),\boldsymbol{g}(u_j)))},
\end{equation}
where $f(\cdot)$ is a dot product to measure the similarity between two augmented user representations.

Our final model combines both the ranking loss and the contrastive loss during training:
\begin{equation}
	\mathcal{L}_{total}=\mathcal{L}_{rank}+\lambda\cdot\mathcal{L}_{cl},
\end{equation}
where $\lambda$ is a hyper-parameter to control the weight of contrastive loss. We will discuss hyper-parameters  of $\lambda$, $p$ and $l$ in subsection \ref{subsection-hyper-parameters-for-contrastiv-learning}.

\subsection{Algorithm Analysis}\label{algorithm-analysisn-subsection}

To make the proposed AW-MoE model more concise and understandable, we summarize it as an algorithm (Algorithm \ref{algorithm:aw-moe}). Given input features including the user behavior sequence $B$, the target item $t$, the query $q$, and other features $o$, AW-MoE outputs the predicted ranking score $\hat{y}$. In summary, the forward propagation of AW-MoE contains four steps. The first step is to compute the input network (line 2-8), within which the main computation is to compute the user representation. The second step is to compute $K$ expert networks (line 9-12). The third step is to compute the gate network (line 13-24), which involves computations for $M$ activation units and $M$ gate units. The fourth step is to compute the overall output, which is the predicted ranking score $\hat{y}$ (line 25-29).

Suppose time complexities of the activation unit $\Phi(\cdot)$, the gate unit $\Theta(\cdot)$, and the expert network $\Psi(\cdot)$ are $\Phi_t$, $\Theta_t$, and $\Psi_t$, respectively. Then the time complexity of AW-MoE is $O(M\Phi_t+K\Psi_t+M(\Theta_t+\Phi_t)+KM+K)$, which can be simplified as $O(M(\Theta_t+\Phi_t)+\Psi_t)$ because $K$ is usually a small constant value.  Suppose space complexities of $\Phi(\cdot)$, $\Theta(\cdot)$, and $\Psi(\cdot)$ are $\Phi_s$, $\Theta_s$, and $\Psi_s$, respectively. Then the space complexity of AW-MoE is $O(\Phi_t+K\Psi_t+\Theta_t+\Phi_t)+O(Vd)$, where $V$ is the vocabulary size of all items and $d$ is the embedding dimension. Since $K$ is usually a small constant value, the space complexity of AW-MoE can be simplified as $O(Vd+\Phi_t+\Psi_t+\Theta_t)$. Both time and space complexities of $\Phi(\cdot)$, $\Theta(\cdot)$, and $\Psi(\cdot)$ are related to the MLP parameters as shown in Figure \ref{the-architecture-of-activation-gate-expert}. As the contrastive learning strategy generates only constant positive instances, and uses in-batch examples as negative instances, it does not increase the time and space complexities of AW-MoE.

\begin{algorithm}[h]
	\caption{The forward propagation of AW-MoE}
	\label{algorithm:aw-moe}
	\LinesNumbered
	\KwIn {$B=\{b_1,...b_j,...,b_M\}$ is the item set of the user behavior sequence, and $M$ is the sequence length. $t$ is the target item. $q$ is the query. $o$ represents other features. $K$ is the number of expert networks.}
	Lookup embeddings and obtain $\boldsymbol{e}_\tau, \tau\in\{b_j,t,q,o\}$\;
	// Step 1: computation for the input network\;
	$\boldsymbol{h}_\tau^I={\rm MLP}^I(\boldsymbol{e}_\tau), \tau\in\{b_j,t,q,o\}$\;
	$\boldsymbol{v}_u^I=\boldsymbol{0}$\;
	\For{j = 1 to M}{
		$\boldsymbol{v}_u^I=\boldsymbol{v}_u^I+\Phi^I(\boldsymbol{h}_{b_j}^I,\boldsymbol{h}_t^I)\boldsymbol{h}_{b_j}^I$\;
	}
	$\boldsymbol{v}_{imp}=\boldsymbol{v}_u^I||\boldsymbol{h}_t^I||\boldsymbol{h}_q^I||\boldsymbol{h}_o^I$\;
	// Step 2: computation for expert networks\;
	\For{k = 1 to K}{
		$s_k=\Psi_k(\boldsymbol{v}_{imp})$\;
	}
	// Step 3: computation for the gate network\;
	$\boldsymbol{h}_\tau^G={\rm MLP}^G(\boldsymbol{e}_\tau), \tau\in\{b_j,q\}$\;
	\For{j = 1 to M}{
		$\boldsymbol{a}_j=\Theta(\boldsymbol{h}_{b_j}^G,\boldsymbol{h}_q^G)$\;
		$w_j^G=\Phi^G(\boldsymbol{h}_{b_j}^G,\boldsymbol{h}_q^G)$\;
	}
	\For{k = 1 to K}{
		$g_k=0$\;
		\For{j = 1 to M}{
			$g_k=g_k+w_j^Ga_{jk}$\;
		}
	}
	// Step 4: computation for the output\;
	$\hat{y}=0$\;
	\For{k = 1 to K}{
		$\hat{y}=\hat{y}+g_ks_k$\;
	}
	\KwOut {$\hat{y}$}
\end{algorithm}


\subsection{System Deployment}\label{system-deployment-subsection}
The AW-MoE has been deployed in the product search engine of JD.com, the largest B2C e-commerce website in China. The online model system and its relation to offline training are summarized in Figure \ref{the-online-model-system}.

\subsubsection{Offline training}

The AW-MoE is trained in an end-to-end manner, with a billion scale dataset collected from search logs of JD.com in the last 15 days. In our initial design of the AW-MoE, the gate network takes the user behavior sequence and the target item as inputs. As a result, the gate network has to be computed for every target item within a search session. Since a gate network computation involves computations of a gate unit and an activation unit for every item in the user behavior sequence, this easily becomes the time-consuming step in both training and inference because of the long user behavior sequence (sometimes over 1,000 items).

In order to train and serve the model efficiently, we aim to reduce the overhead from the gate network. We find that by only using the user and query level features in the gate network, we reduce the number of gate network computations to only once for all target items in a session, and still maintain the prediction accuracy. Our ﬁnal launched model in the search scenario combines the optimization technique above, which results in $>$ 10x saving in computational resource and latency reduction. Nevertheless, in the recommendation scenario, we still use the target item as the input to the gate network due to the lack of query.

\subsubsection{Online serving}

\begin{figure}[h]
	\centering
	\includegraphics[width=0.8\linewidth]{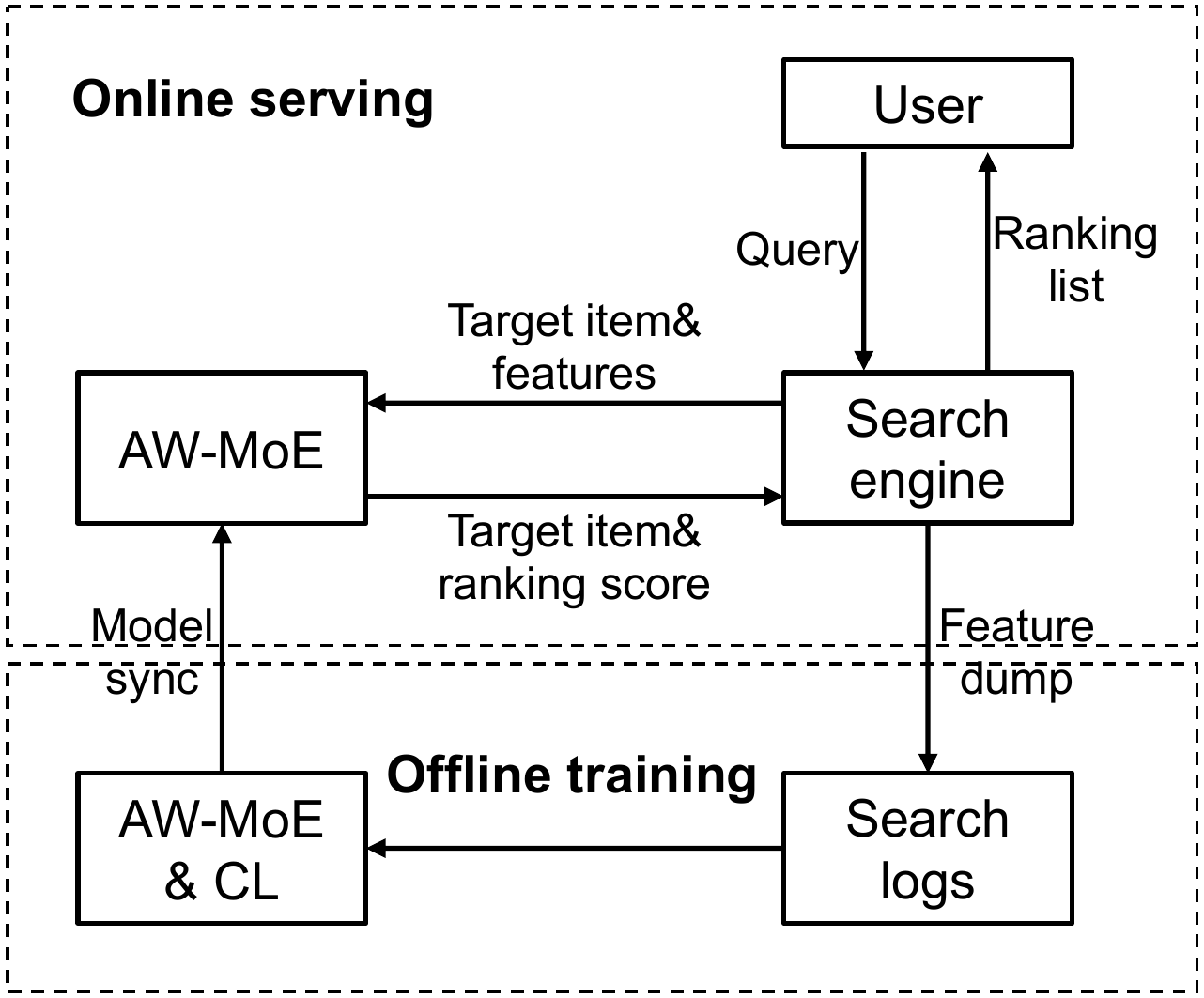}
	\caption{The architecture of our search engine.}
	\label{the-online-model-system}
\end{figure}

The online serving system consists of two components as shown in Figure \ref{the-online-model-system}. The search engine is responsible for distributing and receiving features, and the AW-MoE is responsible for ranking. When a user issues a query, the search engine collects the associated user context features including user behavior sequence, query, as well as user profile information, and then retrieves relevant items using multiple index models~\cite{zhang2021joint-Poeem,xia2021searchgcn-SearchGCN}. Next, the retrieved items and associated features are sent to the AW-MoE, which in turn computes the predicted ranking scores for each item, all of which are sent back to the search engine and are finally presented to the user. For network load balancing, queries are parallel distributed to a cluster with thousands of machines running the AW-MoE model, achieving the average latency of AW-MoE around 20 milliseconds, which meet the requirement of our online service.

\section{EXPERIMENTAL RESULTS}\label{experimental-results-section}

In this section, we compare our proposed AW-MoE model with other state-of-the-art ranking models empirically. In subsection \ref{visualization-of-activation-vectors-learned-by-the-gate-network}, we visualize user representations learned by the gate network. Furthermore, we perform an ablation study on two modules of AW-MoE in subsection \ref{ablation-study-for-the-gate-network}, and investigate effects of three hyper-parameters in contrastive learning in subsection \ref{subsection-hyper-parameters-for-contrastiv-learning}. At last in subsection \ref{experiments-on-online-ab-testing}, we conduct the online A/B testing on the real traffic of hundreds of millions of active users on JD.com.

\subsection{Benchmark Datasets}

Two datasets were used to evaluate our model, including one in-house JD dataset and one public Amazon review dataset.

\subsubsection{In-house JD dataset}

We collected order records from the JD e-commerce search engine. For model training, a total of 6,686,451 search sessions were collected. For each session, the purchased items were labeled as positives and an equal number of negatives were sampled from impressed but not purchased items to balance the ratio of positives to negatives, resulting in a total of 13,466,568 training examples. Each training example had a total of 22 categorical and numerical features, and was converted to an input vector with dimension as 549.

For model testing, three test sets were prepared. Firstly, a total of 76,886 search sessions were collected as the full test set. For each session, the purchased items were labeled as positives and all impressed but not purchased items were labeled as negatives, resulting in a total of 875,395 testing examples. For each training and testing example, items clicked in previous seven days or purchased in previous one year were collected as the user behavior sequence. 
 
Secondly, to further validate the effectiveness of the contrastive learning strategy, two long-tail user test sets were selected from the full test set. The first long-tail user test set consisted of long-tail users having limited historical behaviors, and the second long-tail user test set consisted of long-tail users who were elderly. Statistics of the in-house JD dataset are summarized in Table \ref{tab:jd-data}.

\begin{table}
	\centering
	\caption{Statistics of the in-house JD dataset.}
	\label{tab:jd-data}
	\color{black}
	\begin{tabular}{lrrrrr}
		\toprule
		Statistics & \makecell[c]{Training\\ set} & \makecell[c]{Full\\test set}  & \makecell[c]{Long-tail \\ test set 1} & \makecell[c]{Long-tail \\ test set 2} \\
		\midrule
		\# Sessions & 6,686,451 & 76,886 & 4,695 & 17,650\\
		\# Users & 4,901,748 & 75,180 & 4,694 & 17,225\\
		\# Queries & 1,491,499 & 47,560 & 3,892 & 13,866\\
		\hline
		\# Examples & 13,466,568 & 875,395 & 32,090 & 218,507\\
		Pos : Neg & 1 : 1 & 1 : 10 & 1 : 6  & 1 : 13\\
		\hline
		\makecell[l]{\# Examples / \\ \# Sessions}  & 2.0 & 11.4 & 6.8 & 12.3\\
		\bottomrule
	\end{tabular}
\end{table}

\subsubsection{Public Amazon review dataset}

The public Amazon review dataset~\cite{ni2019justifying-Amazon-dataset-2018} is a widely-used benchmark in recommendation. We preprocessed the dataset following the method described in~\cite{xiao2021adversarial-HSC-MoE}, where all review events were innerly joined with the metadata, resulting in a total of 199,298,798 review events with 13,727,767 users and 6,926,608 items. Then, all review events were grouped by user and were organized in chronological order. The task of this dataset was to predict the last reviewed item for each user given historical reviewed items by the user. We randomly sampled 90\% of all users as the training examples and the rest 10\% of users were used as testing examples. For each training and testing example, the last reviewed item was regarded as the positive item and a negative item was randomly sampled from all other items. Features used in this dataset consisted of features in the original dataset and features added in~\cite{xiao2021adversarial-HSC-MoE}. All features of an example were converted to an input vector with dimension as 354.

Please note that there was no query in this dataset, instead the recommendation was made purely based on the user behavior sequence as well as other basic features. Therefore, the query was replaced by the target item in Figure \ref{the-overall-architecture-of-the-ranking-model}c when evaluating AW-MoE on this dataset. Since the core ideas of our proposed AW-MoE model are the user behavior sequence modeling and the contrastive learning in the gate network, this did not present a severe obstacle to the conclusion in this setting.

\subsection{Evaluation Metrics}

Two metrics were used for performance evaluation: AUC (Area Under the ROC Curve) and NDCG (Normalized Discounted Cumulative Gain)~\cite{jarvelin2002cumulated-NDCG}. AUC measures the probability that a randomly sampled positive item has a higher ranking score than a randomly sampled negative one:
\begin{equation}
	AUC=\frac{1}{|P|}\sum_{p\in P}\frac{1}{|D^+_p||D^-_p|}\sum_{i\in D^+_p}\sum_{j\in D^-_p} \mathbb{I}(\hat{y}_{pi}>\hat{y}_{pj}),
\end{equation}
where $P$ is the set of search sessions in the test dataset, $D^+_p$ and $D^-_p$ are sets of items purchased and not purchased respectively in session $p$, $\mathbb{I}(\cdot)$ is an indicator function, and $\hat{y}_{pi}$ and $\hat{y}_{pj}$ are predicted ranking scores for item $i$ and item $j$ respectively in session $p$.

NDCG is another widely-used metric in ranking, which assigns higher weights to the top-ranked items by applying a position-based discount factor to the ranking score, and then normalize the score by its maximal possible value:
\begin{equation}
	NDCG=\frac{1}{|P|}\sum_{p\in P}(\sum_{i=1}^{|\hat{D}_p|}\frac{y_{pi}}{\log_2(i+1)} / \sum_{i=1}^{|D_p|}\frac{y_{pi}}{\log_2(i+1)}),
\end{equation}
where both $\hat{D}_p$ and $D_p$ are the union set of $D^+_p$ and $D^-_p$, and the difference between them lies in that items in $\hat{D}_p$ are ranked by the predicted score $\hat{y}_{pi}$ while items in $D_p$ are ranked by the ground-truth label $y_{pi}$. 

A higher AUC or NDCG indicates a better performance. Since most users browse only the top-ranked items, we also report AUC and NDCG calculated by the top-10 items in a session, i.e., the AUC@10 and the NDCG@10.

\subsection{Compared Algorithms}

Three state-of-the-art personalized ranking models were compared with our proposed AW-MoE:
\begin{itemize}
	\item \textbf{DNN~\cite{covington2016deep-YouTube-DNN}}. The user representation vector was obtained by sum pooling all items in the user behavior sequence, and was fed into an FFN together with other features for feature interaction.
	\item \textbf{DIN~\cite{zhou2018deep-DIN}}. Similar to DNN except that the user representation vector was computed using the attention mechanism.
	\item \textbf{Category-MoE~\cite{xiao2021adversarial-HSC-MoE}}. It leveraged the MoE framework to learn a ranking model that specialized for each query category by feeding the category id into the gate network.
	\item \textbf{AW-MoE}. Our proposed model that simultaneously fed the user behavior sequence into expert networks and the gate network. Within the gate network, a gate unit and an activation unit were designed for each item in the user behavior sequence to adaptively learn the fine-grained impact of each item to each expert, and the final output of the gate network was obtained by weighted sum of outputs from all gate units using an attention mechanism.
	\item \textbf{AW-MoE \& CL}. Based on the AW-MoE, an auxiliary contrastive loss was imposed to the output of the gate network to improve the robustness of the model and the model generalization for long-tail users.
\end{itemize}

Please note that we did not compare AW-MoE with models like DIEN~\cite{zhou2019deep-DIEN} and DSIN~\cite{feng2019deep-DSIN}, because these models together with DIN~\cite{zhou2018deep-DIN} all were not based on the MoE framework, and all of them exploited the user behavior sequence only once for user representation learning. Therefore, all of these models could be regarded as one expert network without the gate network. On the contrary, our proposed AW-MoE was based on the MoE framework, and mainly focused on the optimization for the gate network, which was designed to learn personalized patterns of feature interactions. Theoretically, all of the aforementioned models were compatible with AW-MoE, and could be the expert network of AW-MoE. As a result, we only compared AW-MoE with the classic DIN model.

\subsection{Parameter Settings}

For experiments on the in-house JD dataset, the FFN networks used in DNN and DIN models had the same network structure as the expert networks used in all MoE-based models, and parameters were illustrated in Figure \ref{the-architecture-of-activation-gate-expert}b. ReLU was used as the activation function for all hidden layers. For all MoE-based models, the number of expert networks $K$ was set as 4. For contrastive learning when training AW-MoE, the masking probability $p$ was set as 0.1, the number of negative examples $l$ was set as 3, and the contrastive loss weight $\lambda$ was set as 0.05, and we investigated effects of these hyper-parameters in subsection \ref{subsection-hyper-parameters-for-contrastiv-learning}. All models were trained on 4 NVIDIA P40 GPUs with  batch size as  1,024. We applied the AdamW~\cite{loshchilov2017decoupled-AdamW} optimizer with an initial learning rate as 1E-4. For experiments on the public Amazon review dataset, we followed the parameter settings (FFN network structure and batch size) as described in~\cite{liu2020kalman-KFAtt}.

\subsection{Performance Evaluation}
\subsubsection{Results From Model Comparison on the JD Dataset}

Evaluation results of five ranking models on the full test set of the in-house JD dataset are illustrated in Table \ref{tab:results-on-the-in-house-jd-dataset-full}. First, our proposed AW-MoE achieved an AUC of 0.8459, which was 2.58\% (3.39\% in terms of NDCG) absolutely higher than DNN. When test items were limited to top-10, improvements were even bigger with 3.26\% and 3.60\% in terms of AUC@10 and NDCG@10, respectively. Second, all MoE-based models outperformed DIN, which is the state-of-the-art model applied in many industrial companies, showing that the MoE framework was consistently beneficial to the personalized ranking. Third, AW-MoE improved Category-MoE, our previous online model, with an absolute AUC gain of 0.71\% (0.99\% in terms of NDCG), which is significant for our business, demonstrating that the user behavior sequence was more suitable than the category id for expert network activation. At last, the contrastive learning strategy (AW-MoE \& CL) further improved the AUC of AW-MoE by 0.13\%, indicating the effectiveness of contrastive learning.

To further validate the effectiveness of the contrastive learning strategy, five ranking models were also evaluated on the two long-tail user test sets (Tables \ref{tab:results-on-the-in-house-jd-dataset-long-tail} and \ref{tab:results-on-the-in-house-jd-dataset-long-tail2}). Take the first long-tail user test set as an example (Table \ref{tab:results-on-the-in-house-jd-dataset-long-tail}), as long-tail users had few historical behaviors, existing ranking models, such as DNN, DIN, and Category-MoE, failed to learn appropriate user representations from user behavior sequence due to data sparsity, hence achieved similar and relatively low AUC values. For the proposed AW-MoE, although it achieved a remarkably high AUC of 0.8353, which was 0.54\% absolutely higher than that of Category-MoE, p-values indicated that the improvement was not significant, showing the difficulty of personalized ranking for long-tail users. Nevertheless, with the contrastive learning strategy, AW-MoE \& CL reached an AUC of 0.8379, which was 0.80\% absolutely higher than that of Category-MoE, and the improvement was significant (p-value $<$ 0.05), demonstrating the effectiveness of the contrastive learning strategy. Furthermore, AW-MoE \& CL improved the AUC of AW-MoE by 0.26\%, which was higher than that on the full test set (0.13\%). When test items were limited to top-10, the improvement was even bigger with 0.49\% in terms of AUC@10, indicating that the contrastive learning strategy did improve the model generalization for long-tail users. Using the second long-tail user test set, similar results were obtained (Table \ref{tab:results-on-the-in-house-jd-dataset-long-tail2}).

\begin{table*}
	\centering
	\caption{Results on the full test set of the in-house JD dataset. Numbers marked with * and ‡ are p-values relative to DNN and Category-MoE, respectively.}
	\label{tab:results-on-the-in-house-jd-dataset-full}
	\begin{tabular}{ccccc|cccc}
		\toprule
		\multirow{2}{*}{Model} & \multicolumn{4}{c|}{Metric}     & \multicolumn{4}{c}{p-value}               \\ \cline{2-9} 
		& AUC    & AUC@10 & NDCG & NDCG@10 & AUC             & AUC@10 & NDCG & NDCG@10 \\
		\midrule
		DNN                    & 0.8201 & 0.7471 & 0.6580 & 0.6369  &-               &-      &-    &-       \\
		DIN                    & 0.8361 & 0.7674 & 0.6785 & 0.6587 & 1.00E-20* &   1.00E-20*    & 1.00E-20*     &  1.00E-20*       \\
		Category-MoE           & 0.8388 & 0.7706 & 0.6820 & 0.6624 & 1.00E-20* &    1.00E-20*    & 1.00E-20*     &   1.00E-20*      \\
		AW-MoE  & 0.8459 & 0.7797 & 0.6919 & 0.6729 & 1.33E-15‡ &   4.38E-10‡     & 1.07E-11‡     & 9.96E-11‡        \\
		AW-MoE \& CL  & \textbf{0.8472} & \textbf{0.7808} & \textbf{0.6937} & \textbf{0.6747} & 1.00E-20‡ &  2.08E-12‡     & 6.66E-16‡     & 5.37E-14‡        \\
		\bottomrule
	\end{tabular}
\end{table*}

\begin{table*}
	\centering
	\caption{Results on the long-tail test set 1 of the in-house JD dataset. Numbers marked with * and ‡ are p-values relative to DNN and Category-MoE, respectively.}
	\label{tab:results-on-the-in-house-jd-dataset-long-tail}
	\begin{tabular}{ccccc|cccc}
	\toprule
	\multirow{2}{*}{Model} & \multicolumn{4}{c|}{Metric}     & \multicolumn{4}{c}{p-value}               \\ \cline{2-9} 
	& AUC    & AUC@10 & NDCG & NDCG@10 & AUC             & AUC@10 & NDCG & NDCG@10 \\
	\midrule
	DNN                    & 0.8274 & 0.7949 & 0.6894 & 0.6802  &-               &-      &-    &-       \\
	DIN                    & 0.8283 & 0.7951 & 0.6909 & 0.6808 & 8.02E-01* &   9.66E-01*    & 7.79E-01*     &  9.07E-01*       \\
	Category-MoE           & 0.8299 & 0.7969 & 0.6916 & 0.6819 & 5.06E-01* &    6.93E-01*    & 6.86E-01*     &   7.68E-01*      \\
	AW-MoE  & 0.8353 & 0.8019 & 0.7004 & 0.6901 & 1.38E-01‡ &   3.09E-01‡     & 1.13E-01‡     & 1.72E-01‡        \\
	AW-MoE \& CL  & \textbf{0.8379} & \textbf{0.8068} & \textbf{0.7034} & \textbf{0.6943} & 2.67E-02‡ &  4.37E-02‡     & 3.34E-02‡     & 3.77E-02‡        \\
	\bottomrule
\end{tabular}
\end{table*}

\begin{table*}
	\centering
	\caption{Results on the long-tail test set 2 of the in-house JD dataset. Numbers marked with * and ‡ are p-values relative to DNN and Category-MoE, respectively.}
	\label{tab:results-on-the-in-house-jd-dataset-long-tail2}
	\begin{tabular}{ccccc|cccc}
		\toprule
		\multirow{2}{*}{Model} & \multicolumn{4}{c|}{Metric}     & \multicolumn{4}{c}{p-value}               \\ \cline{2-9} 
		& AUC    & AUC@10 & NDCG & NDCG@10 & AUC             & AUC@10 & NDCG & NDCG@10 \\
		\midrule
		DNN                    &0.7621  &0.6870  &  0.6039& 0.5820  &-               &-      &-    &-       \\
		DIN                    & 0.7761 &0.7066 &0.6216  & 0.6018&1.24E-06* & 8.62E-08*    & 2.61E-07*     & 1.10E-07*       \\
		Category-MoE           & 0.7772 &0.7059 & 0.6228& 0.6026 & 1.89E-07* &  2.82E-07*    &3.76E-08*     & 3.60E-08*      \\
		AW-MoE  &  0.7849 &0.7151  & 0.6342  & 0.6141 &  7.00E-03‡ &   1.20E-02‡     & 1.00E-03‡     &  2.00E-03‡        \\
		AW-MoE \& CL  & \textbf{0.7873} & \textbf{0.7185} & \textbf{0.6363} & \textbf{0.6166} & 4.00E-04‡ &  5.00E-04‡     & 1.00E-04‡     & 2.00E-04‡        \\
		\bottomrule
	\end{tabular}
\end{table*}

\subsubsection{Results From Model Comparison on the Amazon Dataset}

We further conducted experiments on the public Amazon review dataset. Since the Amazon dataset was a dataset for recommendation, and only one negative example was sampled for each positive example, only the overall AUC metric was calculated for comparison. As shown in Table \ref{tab:results-on-the-public-Amazon-review-dataset}, the conclusions drawn from the Amazon dataset were in agreement with those drawn from the JD dataset. Our proposed AW-MoE achieved an AUC of 0.7362, which was 2.39\%, 2.00\% and 1.09\% absolutely higher than DNN, DIN and Category-MoE, respectively, demonstrating the superiority of the architecture of AW-MoE. The performance of AW-MoE was further enhanced by the contrastive learning strategy (AW-MoE \& CL), which improved the AUC of AW-MoE by 0.19\%. This evaluation again illustrated that our proposed AW-MoE together with the contrastive learning strategy were superior in personalized ranking even on a recommendation dataset, demonstrating the generalization of our approach.

\begin{table}[]
	\centering
	\caption{Results on the public Amazon review dataset. Numbers marked with * and ‡ are p-values relative to DNN and Category-MoE, respectively.}
	\label{tab:results-on-the-public-Amazon-review-dataset}
	\begin{tabular}{ccc}
		\toprule
		Model        & AUC & p-value \\
		\midrule
		DNN          &  0.7123   &   -      \\
		DIN          &  0.7162   &   7.68E-06*     \\
		Category-MoE &  0.7253   &   1.00E-20*    \\
		AW-MoE&0.7362&1.00E-20‡     \\
		AW-MoE \& CL&\textbf{0.7381}&1.00E-20‡     \\
		\bottomrule
	\end{tabular}
\end{table}

\subsection{Visualization of User Representations Learned by the Gate Network}\label{visualization-of-activation-vectors-learned-by-the-gate-network}

\begin{figure}[h]
	\centering
	\includegraphics[width=0.9\linewidth]{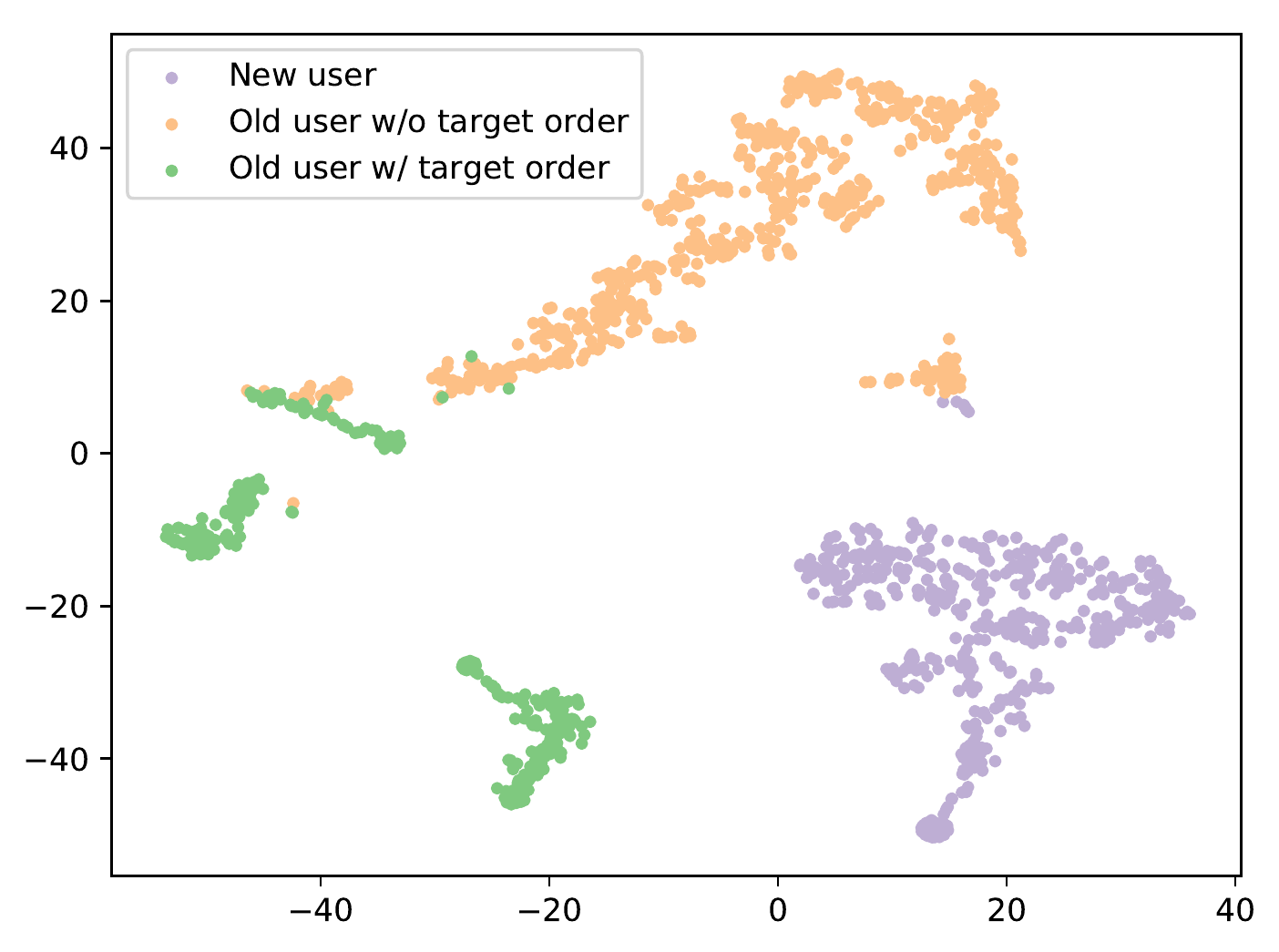}
	\caption{t-SNE~\cite{JMLR:v9:vandermaaten08a-tSNE} visualization of user representations learned by the gate network for different user groups. Users who had no historical behaviors were classified to be new users, otherwise old users. Old users were further classified to users who purchased the target item in the past (Old user w/ target order) and users who did not (Old user w/o target order).}
	\label{the-visualization-of-the-gate-vectors}
\end{figure}

One of the key ideas of the proposed AW-MoE was that we designed a gate network to adaptively activate different expert networks according to the user behavior sequence. As a result, the output of the gate network could be regarded as the user representation. To validate the effectiveness of the proposed model on the in-house JD dataset, we visualized clusters of the user representations learned by the gate network for different groups of users in two-dimensional space using t-SNE~\cite{JMLR:v9:vandermaaten08a-tSNE} as shown in Figure \ref{the-visualization-of-the-gate-vectors}.

As shown in Figure \ref{the-visualization-of-the-gate-vectors}, users who had no historical behaviors (new users) were well clustered and were separated from old users. Moreover, old users were further grouped into users who purchased (Old user w/ target order) and did not purchase (Old user w/o target order) the target item. This observation was in line with the intuition that new users had no prior knowledge to the website and were supposed to activate different experts for feature interactions than the old users. Since old users with target order (green points in Figure \ref{the-visualization-of-the-gate-vectors}) usually had abundant historical behaviors, their shopping interests were more diverse, hence these users were further grouped into several sub-clusters, whose clustering rules were too complicated to be explained. The visualization results indicated that different groups of users did have different representations, and thus activated different expert networks, which was able to improve the learning effectiveness for each expert and hence led to a better personalized ranking performance.

\subsection{Ablation Study for the Gate Network}\label{ablation-study-for-the-gate-network}

In this subsection, we performed an ablation study on AW-MoE by showing how the gate unit (GU) and the activation unit (AU) affected its performance. Within the gate network of AW-MoE, both the gate unit and the activation unit were designed to model the user behavior sequence, but their motivations were different. The gate unit was to learn the effect of the item to experts, while the activation unit was to learn the importance of the item. 

As shown in Table \ref{tab:results-of-the-ablation-study-for-the-gate-network}, if the output of the gate network was calculated by sum pooling all items in the user behavior sequence without the gate unit and the activation unit (Base), it achieved an AUC of 0.8438 and an NDCG of 0.6884, which were fairly high compared with category-MoE in Table \ref{tab:results-on-the-in-house-jd-dataset-full}, indicating that the user behavior sequence was more suitable than the category id for expert network activation. If either the gate unit or the activation unit was added, a more fine-grained user representation was learned by the gate network, hence the AUC and the NDCG were improved by around 0.1\% and 0.2\%, respectively. Furthermore, if both the gate unit and the activation were added, AW-MoE achieved an  AUC of 0.8459 and an  NDCG of 0.6919, which were 0.21\% and 0.35\% absolutely higher than those of the base model, demonstrating that both the gate unit and the activation unit had unique values, and combining two modules together achieved the best performance.

\begin{table}[]
	\centering
	\caption{ Performance comparison of AW-MoE on the full test set of the in-house JD dataset w.r.t. different modules in the gate network. GU: gate unit, AU: activation unit.}
	\label{tab:results-of-the-ablation-study-for-the-gate-network}
	\begin{tabular}{ccccc}
		\toprule
		Model        & AUC & AUC@10 & NDCG & NDCG@10\\
		\midrule
		\makecell[c]{Base (sum pooling\\ of behaviors)}          &  0.8438  &   0.7766 & 0.6884 & 0.6692      \\
		Base+GU          &  0.8451   &   0.7784 & 0.6900 & 0.6709    \\
		Base+AU &  0.8455  &  0.7789 & 0.6908 & 0.6717   \\
		\makecell[c]{Base+GU+AU\\ (AW-MoE)} & \textbf{0.8459} & \textbf{0.7797} & \textbf{0.6919} & \textbf{0.6729}     \\
		\bottomrule
	\end{tabular}
\end{table}

\subsection{Hyper-Parameters for Contrastive Learning}\label{subsection-hyper-parameters-for-contrastiv-learning}

In this subsection, we investigated effects of hyper-parameters used in the contrastive learning strategy, including the mask probability $p$, the number of negative instances $l$, and the weight for contrastive loss $\lambda$. As the contrastive learning strategy was proposed mainly for long-tail users and AUC@10 was more sensitive to the performance change (see Table \ref{tab:results-on-the-in-house-jd-dataset-long-tail}), we performed the comparison on the long-tail user test set of the in-house JD dataset and tuned hyper-parameters according to the metric of AUC@10.

\begin{figure*}[h]
	\centering
	\includegraphics[width=0.95\linewidth]{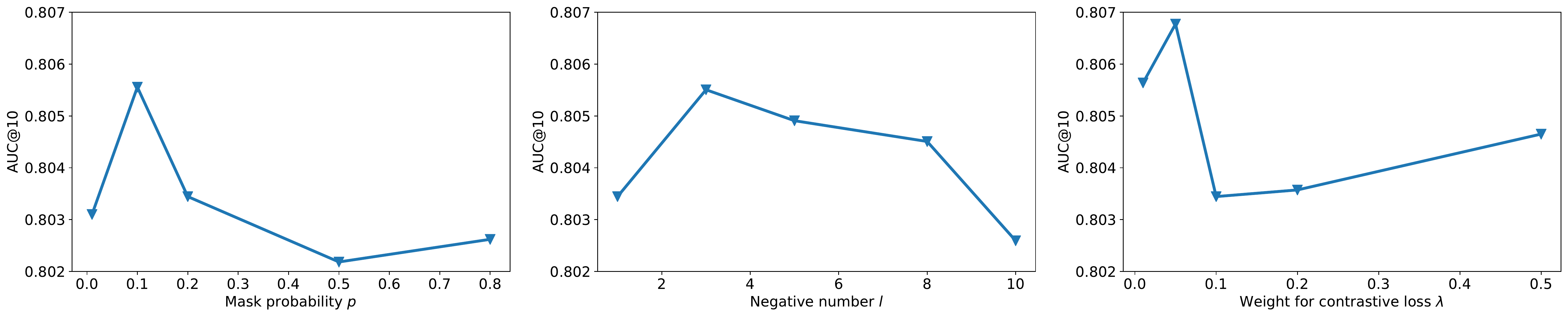}
	\caption{Performance comparison of the contrastive learning strategy on the long-tail user test set of the in-house JD dataset w.r.t. different $p$, $l$, and $\lambda$. First, $p$, $l$, and $\lambda$ were initialized as 0.2, 1, and 0.05, respectively. Then, $l$ and $\lambda$ were fixed, and $p$ was tuned and was found to be optimal at 0.1. Next, $p$ was revised as 0.1, and $l$ was tuned and was found to be optimal at 3. At last, $l$ was revised as 3, and $\lambda$ was tuned and was found to be optimal at 0.05.}
	\label{fig-hyper-parameters-for-contrastiv-learning}
\end{figure*}

As shown in Figure \ref{fig-hyper-parameters-for-contrastiv-learning}, a general pattern was found that the performance peaked when hyper-parameters were set at a special value and then deteriorated if hyper-parameters were increased or decreased. Take the mask probability $p$ as an example, when it was set as a very small value (0.01), the mask operator can hardly work, and when it was set as values greater than 0.2, too many items were masked thus hurting the performance. Therefore, the optimal $p$ was set as 0.1. Similarly, the optimal $l$ and $\lambda$ were set as 3 and 0.05 respectively.

\subsection{Experiment on Online A/B Testing}\label{experiments-on-online-ab-testing}

From 2021-Sep-17 to 2021-Sep-22, online A/B testing was conducted in the JD e-commerce search engine. AW-MoE contributed 0.78\% (p-value=2.20E-5) user conversation rate (UCVR) and 0.35\% (p-value=2.97E-5) user click through rate (UCTR) gain compared with the previous Category-MoE model online. Please note that every 0.10\% increase in UCVR or UCTR brings great revenues for the company, hence the improvement achieved by the AW-MoE was significant. As a result, the proposed AW-MoE ranking model has been deployed in the JD e-commerce search engine.

\section{Conclusion and future work}

We presented Attention Weighted MoE with contrastive learning for personalized ranking in e-commerce. AW-MoE leveraged multiple expert networks to capture the diverse feature interaction patterns, and the attention weighted gate network to explicitly learn the personalized experts activation vectors. In particular, the gate network incorporated a gate unit and an activation unit for each item in the user behavior sequence to learn fine-grained activation schemas for expert predictions. Furthermore, an auxiliary contrastive loss was imposed to the output of the gate network to improve both the robustness of the model and the generalization for long-tail users. Our experiments demonstrated that AW-MoE achieved significant improvement in personalized ranking on both search and recommendation datasets. AW-MoE has now been deployed in the JD e-commerce search engine, serving the real traffic of hundreds of millions of active users.

For future works, we plan to update the vanilla MoE to the sparsely-gated MoE~\cite{shazeer2017outrageously-Top-K-MoE} by increasing the number of experts and introducing a Top-K gate network. Furthermore, the adversarial regularization used in~\cite{xiao2021adversarial-HSC-MoE} is a promising technique to encourage the disagreement among different experts, thus improving the diversity of perspectives in the final ensemble. At last, more data augmentation strategies for the user behavior sequences will be explored, such as item reordering~\cite{xie2022contrastive,zhu2021contrastive}. to further improve the robustness and generalization of the model.

\bibliographystyle{IEEEtran}
\bibliography{reference}

\end{document}